\documentclass[useAMS,usenatbib]{mn2e}
\usepackage{graphicx}
\usepackage{url}
\usepackage{indentfirst}
\newcommand{\lsim} {\mathrel{\hbox{\rlap{\lower.55ex \hbox{$\sim$}}
                             \kern-.3em \raise.4ex \hbox{$<$}}}} 

\newcommand{\gadget}{{\scshape gadget}-2 }
 
\title[Free-free signal in clusters]{The cosmological free-free signal from galaxy groups and clusters}
\author[Ponente et al.]{P.P. Ponente$^{1,2}$, J.M. Diego$^1$, R.K. Sheth$^3$, 
                        C. Burigana$^4$, S.R. Knollmann$^5$, Y. Ascasibar$^5$\\
			$^1$ IFCA, Instituto de F\'\i sica de Cantabria (UC-CSIC),  
			Av. de Los Castros s/n, 39005 Santander, Spain\\
   			$^2$ Departamento de F\'\i sica Moderna, Av. de Los Castros s/n, 39005 Santander, Spain\\ 
			$^3$ Center for Particle Cosmology,
			University of Pennsylvania 209 S. 33rd Street, Philadelphia, PA 19104-6396, USA.\\
			$^4$ INAF/IASF, Istituto di Astrofisica Spaziale e Fisica Cosmica di Bologna,
			via Gobetti 101, 40129, Bologna, Italy.\\
			$^5$ Grupo de Astrof\'\i sica, Departamento de F\'\i sica Teorica, 
			Universidad Aut\'onoma de Madrid, Cantoblanco, 28049, Spain.}

\begin{document} 
 
\date{Draft version \today} 
 
\pagerange{\pageref{firstpage}--\pageref{lastpage}}
 
\maketitle 
 
\label{firstpage} 
 
%%%%%%%%%%%%%%%%%%%%%%%%%%%%%%
\begin{abstract} 
%%%%%%%%%%%%%%%%%%%%%%%%%%%%%%

Using analytical models and cosmological N-body simulations, we study the free-free radio emission from ionized gas in clusters and groups of galaxies. The results obtained with the simulations are compared with analytical predictions based on the mass function and scaling relations. Earlier works based on analytical models have shown that the average free-free signal from small haloes (galaxies) during and after the reionization time could be detected with future experiments as a distortion of the CMB spectrum at low frequencies ($\nu <$ 5 GHz). We focus on the period after the reionization time (from redshift $z=0$ up to $z=7$) and on haloes that are more massive than in previous works (groups and clusters). We show how the average signal from haloes with $M > 10^{13} h^{-1} M_{\odot}$ is less than 10\% the signal from the more abundant and colder smaller mass haloes. However, the individual signal from the massive haloes could be detected with future experiments opening the door for a new window to study the intracluster medium. 

\end{abstract}

%%%%%%%%%%%%%%%%%%%%%%%%%%%%%%
\section{Introduction}
%%%%%%%%%%%%%%%%%%%%%%%%%%%%%%
 
In the near future, new high sensitivity experiments observing at radio and millimeter wavelengths will open a new window to study the high redshift Universe and in particular the re-ionization period. Among these experiments, the {\it Square Kilometer Array} (or SKA hereafter, \citealp{SKA_science_overview}) and the {\it Atacama Large Millimeter Array} (or ALMA hereafter, \citealp{ALMA_project}) are the most relevant ones due to their sensitivity and angular resolution. 
These experiments will be able, for the first time, to trace in detail the distribution of neutral hydrogen before re-ionization (through the 21 cm line, see e.g. \citealp{Schneider_et_al_2008}), and the transition between a neutral and ionized Universe at the time of reionization (ALMA could see the first galaxies emerging at the reionization time).\\

The study of the re-ionization period will offer a unique window to help us understand the formation of the first stars and galaxies. 
The possibilities of this new window for astronomy has motivated many studies that focus, for instance, on the 21-cm line radiation from neutral gas (\citealp{Scott_Rees_1990, Miralda_Escude_Rees_1998, Kunth_et_al_1998, Oh_Mack, Tozzi_et_al_2000, Ciardi_Madau_2003, Furlanetto_et_al_2004, Zaldarriaga_et_al_2004, Burigana_et_al_2004, Gnedin_Prada_2004}), or on the kinematic Sunyaev-Zeldovich effects (kSZ, \citealp{kSZ_1980}) from inhomogeneous (patchy) reionization on large scales (\citealp{Santos_et_al_2003, Iliev_et_al_2007, Jelic_et_al_2010}). In \cite{Peng_Oh_1999}, it is proposed that the reionization can be studied also through the H-$\alpha$ emission, useful to trace young star formation regions.  
%Complementary to the radio observations of the 21 cm line of neutral hydrogen (\citealp{}), the newly ionized regions could be {\bf also} detected through their H$-\alpha$ emission (\citealp{Peng_Oh_1999}), or their interaction with the cosmic microwave background (or CMB) photons ({\bf kinematic} Sunyaev-Zel'dovich or kSZ effect, \citealp{SZ_1972}). 

Another signal emerging from the ionized regions will be the free-free from interactions between the electrons and ions in the plasma. The photons emerging from these interactions can be observed in the radio and microwave bands.~The distortion that free-free induces on the background temperature in the Rayleigh-Jeans part of the spectrum (\citealp{Bartlett_Stebbins}) is actually constrained by the ground based measurement of \cite{Bersanelli_et_al_1994} at 2~GHz, $Y_{\rm{ff}} < 1.9 \times 10^{-5}$ (95\% CL). 
%The {\small FIRAS} instrument (Far and Infrared Radio Absolute Spectrophotometer, see \citealp{FIRAS}) set a constraint on the optical depth of free electrons on the corresponding average free-free signal, by looking at the distortion of the CMB energy spectrum at low frequencies.  {\bf The constrain (95\% confidence limit) to the Compton $y$ {\bf distortion} comes from {\small FIRAS} and has been given first in \cite{FIRAS} with $y = 2.5 \times 10^{-5}$, then in \cite{Fixsen_et_al_1996} with the actual value of $y=1.5 \times 10^{-5}$}, where $y$ is proportional to the optical depth of the free electrons (see also \citealp{Salvaterra_Burigana_2002}). 

Most of the efforts focus on the study of the 21-cm line and the interaction between the CMB photons and the ionized clouds but little has been done in relation to the free-free signal. In this paper we focus on the free-free emission and its ability to trace the ionized medium. The free-free emission (or $bremsstrahlung$) can be potentially observed in the local Universe and up to the re-ionization era. UV radiation emerging from the first stars and quasars ionized the neutral hydrogen creating expanding bubbles of ionized plasma. During a free-free interaction between two charged particles (free electrons and ionized atoms), the electron loses part of its kinetic energy by emitting a photon. The energy of the photon ranges from the radio to the X-ray wavelength depending on the electron temperature.~Since this interaction involves two particles, its intensity depends on the square of the free electron (or equivalently the ion) density, $n_{\rm{e}}$. This $n_{\rm{e}}^2$ dependence makes the free-free signal an interesting candidate for cross-correlations with other signals like the SZ effect where the signal amplitude depends linearly on $n_{\rm{e}}$.\\ 

In the late 90's, an experiment was designed to measure the distortion of the CMB spectrum due to the cosmological free-free signal; the Absolute Radiometer for Cosmology, Astrophysics and Diffuse Emission (or {\small ARCADE}, see \citealp{Fixsen_et_al_2004,Kogut_et_al_2006,Fixsen_et_al_2009, Seiffert11} for details). Its goal is to detect the average free-free signal at frequencies around 1 GHz. Studying the distortion of the CMB spectrum at these frequencies would allow, in principle, to set strong constraints on the history of reionization of the Universe. 
Recently, the ARCADE team presented the results of the ARCADE2 mission that studies both Galactic and extragalactic signals (\citealp{Fixsen_et_al_2009}).~They detect a signal that is significantly larger than the expected extragalactic radio background (a factor $\sim5$ brighter than the estimated contribution from radio point sources). The ARCADE team is currently exploring the possible causes of such a signal like for instance possible foreground contamination, synchrotron emission from Earth's magnetic field or CII lines. In the latest review of the results of the mission (\citealp{Seiffert11}), the authors still report that the excess detection remains unexplained, even though the three main sources of errors, Galactic emission, instrumental systematic errors and radio emission from the faint end of the distribution of known sources, are carefully taken into account.~\cite{Sharpe_2009} has suggested that the observed excess is produced by synchrotron radiation emerging from an optically thin low density magnetized plasma region in the heliosphere of the Sun.\\ 
 %Based on predictions made in earlier works with analytical models, the ARCADE mission should be able to measure the departure from the blackbody spectrum and hence the average cosmological free-free signal.\\

 \cite{Peng_Oh_1999} presents an exhaustive treatment of the different sources of radiation that could be detected with SKA and ALMA in the range of the radio frequencies. He pays special attention to the free-free signal from small haloes and the intergalactic medium  (or IGM) and concludes that the IGM signal is subdominant when compared with the signal from haloes. Another interesting work is presented in \cite{Cooray_Furlanetto} where the authors use a halo model to predict the amount of free-free signal from haloes.~The authors also compute the angular power spectrum of the signal produced by the free-free below 2 GHz and make predictions in the context of the ARCADE mission.
\cite{Burigana_et_al_1995} discusses different physical processes involving the CMB photons and the ionized medium, including also the free-free signal.  
In \cite{Burigana_et_al_2004} the authors discuss about the possibility of detecting the individual sources of free-free signal. 

All these works have focused on the signal from small and cold haloes, largely ignoring the signal coming from larger and hotter haloes (groups and clusters). In this paper we will study the regime of more massive haloes and focus on the period after reionization. Also, an important advantage of working with more massive haloes is that their modeling is much simpler than in the case of smaller haloes.~The cooling time is significantly larger for massive haloes and one can more easily ignore highly non-linear phenomena like radiative cooling.

\section{Free-Free emission}
%%%%%%%%%%%%%%%%%%%%%%%%%%%%%%

In a hot plasma with temperature $T$, the electrons move with kinetic energy $E_{\rm e} = {3/2k_{\rm b} T}$ where $k_{\rm b}$ is the Boltzmann's constant ($1.38 \times 10^{-23}$ J/K). The minimum $T$ required to ionize a plasma is  $\approx 2 \times 10^4$ K (\citealp{zaldarriaga_hui_tegmark}). This is also the temperature at which most of the cooling radiation occurs in a typical galaxy halo (\citealp{Fardal_et_al}). 
This temperature can be translated into a kinetic energy for the free electrons, typically in the order of $2 \times 10^{12}$ erg ($\sim1$ eV). The collisions between opposite charged particles within the plasma modify the paths of the electrons that lose a few percent of their kinetic energy ({\it bremsstrahlung} or {\it brake radiation}). The net effect is the bulk emission of photons in the radio frequency range ($1 \sim 10$ GHz), strongly dependent on the square of the electron density. Note that this square dependence implies a crucial role of the density contrast pattern inside the haloes.
%%%In a previous work, \citealp{Peng_Oh_1999}, a temperature of $10^4$ K is assumed for all the ionized clouds. \citealp{Peng_Oh_1999} assumes that the free-free emission comes from HII regions around stars, not from gas heated by gravitational collapse. 
%%%This temperature can be translated into a kinetic energy. In the case of electrons it corresponds to $2\times 10^{-12}$ erg ($\sim 1$ eV). When an electron interacts with an ion its path changes and the electron looses energy ({\it bremsstrahlung}) by emitting one photon. The loss is several orders of magnitude smaller than the electron's original energy, $\Delta E_e \approx 4 \times 10^{-5}$ eV. 
%%%Each time an interaction occurs, a photon is emitted at the frequency: $\nu = \Delta E_e / h$. This frequency corresponds to about 1-10 GHz, that is, in the range of the radio-frequencies. Since the interaction involves two particles, the free-free signal is proportional to the product of the densities of the ion and the electron. As an approximation, if one assumes a plasma of ionized Hydrogen, both densities are equivalent and hence the free-free signal is proportional to the square of the density. 

 The bremsstrahlung, or free-free signal, can be parametrized in terms of the electron density $n_{\rm{e}}$ and temperature $T_{\rm{e}}$ as (see for example \citealp{Rybicki_Lightman_1979, Cooray_Furlanetto, Oh_Mack});
\begin{equation}
\epsilon_{\rm \nu} = 5.4 \times 10^{-39} n_{\rm{e}}^2 T_{\rm{e}}^{-1/2} g_{\rm{ff}}(\nu,T_{\rm{e}}) \exp\left( {\frac{-h{\rm{\nu}}}{k_{\rm b} T_{\rm e}}} \right ),
\label{ff_c}
\end{equation}
in units of ergs cm$^{-3}$ s$^{-1}$ Hz$^{-1}$ sr$^{-1}$. The Gaunt factor (\citealp{gaunt_factor, Burigana_et_al_1995}), $g_{\rm ff}$, is computed for the observed frequency but it has a weak dependency on the temperature of the gas. In the Rayleigh-Jeans limit (where the free-free radiation is more relevant) the exponential part can be safely neglected. From Eq. (\ref{ff_c}) it is clear that the free-free emissivity depends mostly on the electron density $n_{\rm{e}}$. The inverse dependence with the root square of temperature is a direct consequence of the thermal Maxwellian distribution.

The luminosity of an ionized volume of space with constant $n_{\rm{e}}$ and $T_{\rm e}$ can be obtained from Eq. (\ref{ff_c}) by integrating the electron density and temperature over that volume.~By dividing this luminosity by the corresponding luminosity distance, the flux (or brightness) in Jy can be derived ($1\mbox{ Jy }= 10^{-23}$ergs/s cm$^2$ Hz). 
In terms of the temperature distortion, the brightness can be transformed into equivalent temperature. In the Rayleigh-Jeans limit we have that 
\begin{equation}
\Delta T \propto F \lambda^2
\label{eq_DeltaT}
\end{equation}
 where $F$ is the free-free flux and $\lambda = c/\nu$. Thus, at 1 GHz, while the flux does not change much with frequency, the free-free temperature distortion is expected to be higher than at 10 GHz by a factor 100. This simple scaling shows the convenience of looking for the free-free signal at lower frequencies. Several attempts have been made in the past to measure the free-free distortion at low frequencies as a deviation of the nearly perfect CMB blackbody energy spectrum. The first accurate measurements of the spectrum of the CMB were made by the FIRAS and have shown no departure from the blackbody spectrum (within the error bars) in the frequency range of $60 - 600$ GHz. 
It is expected that new experiments will detect the average free-free contribution  at lower frequencies in the shape of a distortion of the CMB energy spectrum.\\ 

%%%%%%%%%%%%%%%%%%%%%%%%%%%%%%%%%%%%%%%%%%%%%%%%%%%%
\section{Predictions from analytical models}
%%%%%%%%%%%%%%%%%%%%%%%%%%%%%%%%%%%%%%%%%%%%%%%%%%%%

Through analytical halo models it is possible to explore a wide range of cases. \cite{Peng_Oh_1999} shows that the free-free contribution coming from the diffuse IGM is significantly smaller than the signal from ionized haloes so it can be safely ignored. Two ingredients are needed in order to compute the average free-free signal from haloes. First, the mass function, $n(M,z)$, that predicts the average number of haloes per redshift, $z$, and mass, $M$, intervals and, second, a model for the internal gas distribution (and temperature) inside the haloes. The abundance of haloes can be computed from the mass function given a cosmological model. We use the mass function of \cite{Sheth_Tormen_1999} (or ST mass function hereafter) for this purpose. The ST mass function reproduces well the results obtained with large N-body simulations. For the internal distribution of the gas in the haloes and temperature we assume a standard isothermal $\beta$-model with $\beta=2/3$. The gas density profile plays an important role since steeper profiles can produce a larger free-free signal with the same amount of gas (as it happens in the X-ray band with gas in galaxy clusters). Other more realistic models can be found in the literature (see for instance \citealp{Ascasibar_et_al_2003, ascasibar_diego_2008}) but for simplicity we will use the $\beta$-model as this model requires only three parameters ($T_{\rm o}$ for the temperature and $R_{\rm{c}}$, and $n_{\rm{o}}$ for the $\beta$-model). 

%%%%%%%%%%%%%%%%%%%%%%%%%%%%%%%%%%%%%%%%%%%%%
\subsection{Predictions for a single halo}
%%%%%%%%%%%%%%%%%%%%%%%%%%%%%%%%%%%%%%%%%%%%%

The $\beta$-model is widely used in the context of galaxy clusters to describe the electron density as a function of radius (e.g. \citealp{Cavaliere_fusco_femiano_1976, Diego_Majumdar_2004}):
\begin{equation}
n_{\rm{e}}(R) = \frac{n_{\rm{o}}}{1+\left ( \frac{R}{R_{\rm{c}}}\right )^2},
\label{betamodel}
\end{equation}
where $n_{\rm{o}}$ is the electron density at the center of the halo and $R_{\rm{c}}$ is the core radius and we have assumed $\beta=2/3$.
 
For the $\beta$-model, the free-free luminosity can be computed integrating Eq. (\ref{ff_c}) over the volume of a sphere of radius $R_{\rm vir}$
\begin{eqnarray}
L_\nu & = & \int_V \epsilon_\nu dV  \nonumber \\
 %            & = &  \mbox{C1}\cdot \int_0^{R_{vir}} 4\pi n_{\rm{e}}^2(R) R^2 dR  \\ \nonumber
             & = &  \mbox{C} n_{\rm{o}}^2R_c^3 \left (\tan^{-1}\sqrt{p}- \frac{\sqrt{p}}{p+1}\right ) 
\label{BETA_primitive}
\end{eqnarray}
where $\mbox{C} = 5.4 \times 10^{-39} 2\pi T_{\rm e} ^{-1/2} g_{\rm ff}(\nu,T_{\rm e})$
and the argument $p$ is the ratio $R_{\rm vir}/R_{\rm c}$. For simplicity we have dropped the negligible term $exp(-h\nu/(k_{\rm b} T_{\rm e}))$ in Eq. (\ref{ff_c}).   

The halo luminosity can be transformed into flux given the luminosity distance, $D_{\rm L}$, from the halo at redshift $z$ to the observer (at $z=0$). 
\begin{equation}
S_{\rm \nu}(Jy) = \frac{L_{\rm \nu}}{4\pi D_{\rm L}(z)^2}.
\label{brightness}
\end{equation}

The values of $n_{\rm{o}}$, $R_{\rm{c}}$ and $T$ can be computed from scaling relations.
In order to establish these relationships we assume that the total mass of the halo, $M$, and the total mass of the gas, $M_{\rm gas}$, are proportional to each other with the proportionality constant being the universal baryon fraction, $f_{\rm{b}} = M_{\rm gas}/M$. 
Given a virial mass for the halo, the virial radius can be expressed as (\citealp{Finoguenov_et_al_2001, Verde_et_al_2001}) 
\begin{equation}
R_{\rm vir} = 1.3 M_{15}^{1/3} (1+z)^{-1},
\label{raggioviriale}
\end{equation}     
where $M_{15}$ is the halo total mass expressed in $10^{15} h^{-1}M_\odot$ and the radius is scaled with the expansion factor $(1+z)^{-1}$.~Within the virial radius, the relation between the baryonic mass and the electron density profile (given in Eq. (\ref{betamodel})) is
\begin{eqnarray}
M_{\rm gas} & = &\int_V \mu  m_p n_{\rm{e}}(R)dV \\ \nonumber
        & = & 4 \pi \mu m_{\rm p} \int_0^{R_{\rm vir}} n_{\rm{e}}(R')R'^2dR'.
\label{betamodel2}
\end{eqnarray}

Then, a relationship between $n_{\rm{o}}$ and the total mass of virialized halo, $M_{\rm{v}} \approx M_{\rm gas}/f_{\rm b}$, can be established:
\begin{equation}
n_{\rm o} = \frac{M_{\rm v} f_{\rm b}}{(p - tan^{-1}(p)) 4\pi \mu m_{\rm p} R_{\rm c}^3},
\end{equation}
where $p=R_{\rm vir}/R_{\rm c}$. The ratio between $R_{\rm vir}$ and $R_{\rm c}$ is assumed to be constant ($R_{\rm vir}/R_{\rm c}=10 $). For $f_{\rm b}$ we assume $f_{\rm b} = 0.13$. Finally, for the temperature we use the relation obtained by \cite{Diego_et_al_2001} which was shown to be consistent with X-ray measurements  
\begin{equation}
T(\mbox{keV}) = 10 M_{15}^{4/7}.
\label{Tscalinglaw}
\end{equation}

Once $n_{\rm o}$, $R_{\rm v}$, $R_{\rm c}$ and $T$ are known, it is possible to compute the total free-free luminosity, flux and temperature distortion of the halo at redshift $z$ from equations (\ref{BETA_primitive}), (\ref{brightness}) and (\ref{eq_DeltaT}) respectively.

%%%%%%%%%%%%%%%%%%%%%%%%%%%%%%%%%%%%%%%%%%%%%%%%%%%%%%%%%%%%%%%%%
\subsection{The abundance of haloes: the mass function}
%%%%%%%%%%%%%%%%%%%%%%%%%%%%%%%%%%%%%%%%%%%%%%%%%%%%%%%%%%%%%%%%%

In this work we use Sheth \& Tormen (ST hereafter) mass function (\citealp{Sheth_Tormen_1999}, see Appendix A). 

We compute the mass function between the masses $10^8 \leq M \leq 10^{16} h^{-1}M_\odot$. This mass interval covers the range from the smallest ionized haloes to the largest galaxy clusters. We include the small haloes in our calculation for comparison purposes with earlier works and with the more massive haloes. In our simple model we will make the assumption that all haloes included in the mass function are fully ionized. This is not properly true in the low end of the mass interval since, as it was discussed in \cite{Peng_Oh_1999}, the low mass haloes will stay ionized only for a limited amount of time. Consequently, at a given redshift, only a fraction of the low mass haloes are {\it active} or fully ionized.~The conclusions derived from our calculations should be then considered as an upper limit in the low mass range ($M$ $\lsim$ $10^{12} h^{-1}M_\odot$). On the other hand, the high mass end haloes can be considered as fully ionized as most the gas in these massive haloes (clusters) will remain ionized by the high temperatures of the plasma in the clusters. 
Regarding the redshift range we will consider only the contributions up to redshift $z=7$. The reionization period was studied in \cite{Loeb,Peng_Oh_1999,Oh_Mack}. 
\begin{figure}
\begin{center}
\includegraphics[width=8.0cm]{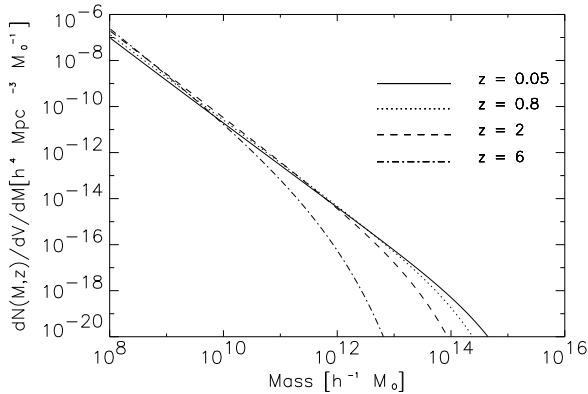}
\end{center}
\caption{Mass function for different redshift intervals. Note how small haloes are common at all redshifts (their population drops at redshifts larger than the ones shown here).}
\label{mass_function_in_masses}
\end{figure}
In Fig. \ref{mass_function_in_masses} we show how the mass function behaves for different redshifts.~The low mass haloes  ($\sim 10^8$ solar masses) show a nearly constant abundance at all redshifts while the number of massive haloes decreases with redshift.\\

In the next subsection we will combine the predicted flux of the $\beta$-model from one halo with the mass function to compute the mean free-free signal from a cosmological volume. 
%%%%%%%%%%%%%%%%%%%%%%%%%%%%%%%%%%%%%%%%%%%%%%%%%%%%%%%%%%%%%%%%%%%%%%%%%%
\subsection{Average free-free emission from an analytical halo model}
%%%%%%%%%%%%%%%%%%%%%%%%%%%%%%%%%%%%%%%%%%%%%%%%%%%%%%%%%%%%%%%%%%%%%%%%%%

\begin{figure}
\begin{center}
\includegraphics[width=8.0cm]{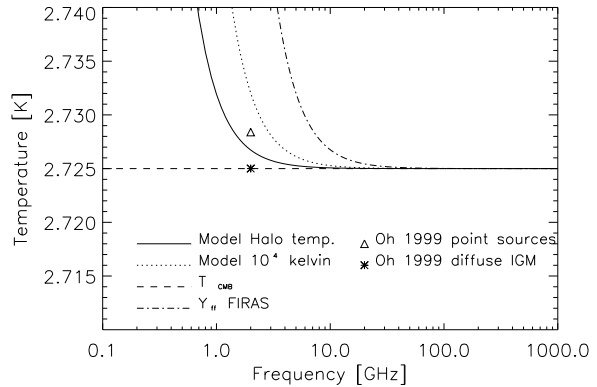}
\end{center}
\caption{  Average temperature distortion due to free-free as a function of frequency for our analytical model. The solid line shows the distortion obtained assuming that the temperature of each halo was computed with the scaling law $T(\mbox{keV}) = 10 M_{15}^{4/7}$; for reference, we show as a  dotted line the distortion corresponding to a fixed temperature of $10^4$ K for all haloes. The dot-dashed line represents the 95\% confidence level observational upper limit derived from \citep{Bersanelli_et_al_1994}. The star and the triangle represent the results from \citep{Peng_Oh_1999} related to the diffuse IGM ($\Delta T=6.0\times 10^{-6}$ K) and to point sources ($\Delta T=3.4\times 10^{-3}$ K) respectively.}
\label{FFmodel}
\end{figure}

Combining a model for the gas distribution inside a halo, like the $\beta$-model, with the abundance of haloes as a function of mass and redshift, it is possible to compute the mean free-free signal in a solid angle as a function of redshift and/or mass. We can also integrate this information in the redshift-mass space and estimate the mean free-free signal from all these haloes. 

Given a redshift and mass interval, we compute the number of haloes in the interval and compute the free-free flux from those haloes. After integrating over the entire redshift range ($0<z<7$) and mass range ($10^8 < M < 10^{16}$) we compute the mean free-free flux from all the haloes.~The flux is converted into thermodynamic temperature to compute the $\Delta T/T$ as a  function of the frequency. The resulting distortion from our analytical model is shown in Fig. \ref{FFmodel}. When comparing our results with those obtained by \cite{Peng_Oh_1999}, we find that our model (solid line) falls below the predicted value by \cite{Peng_Oh_1999}. This can be explained by the fact that we are assuming a higher temperature for the haloes. Fixing the temperature to $T = 10^4$ K (like in \citealp{Peng_Oh_1999}), our model predicts more signal than in \cite{Peng_Oh_1999}. A possible explanation is that, in \cite{Peng_Oh_1999}, only a fraction of the haloes were {\it active} while in our case all haloes are ionized.\\ 

\begin{figure}
\begin{center}
\includegraphics[width=8.0cm]{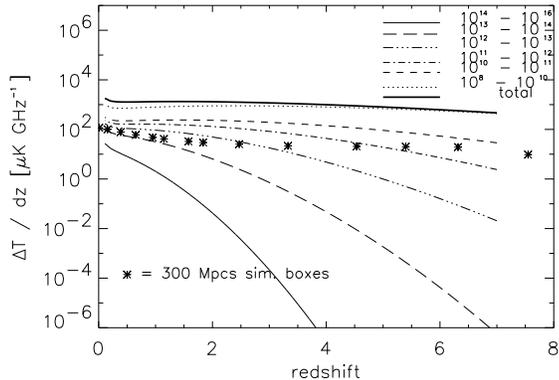}
\end{center}
\caption{Free-free signal for different mass intervals as a function of redshift and for $\nu = 1$ GHz. 
  The points represent the distortion $\Delta T$ computed from a cosmological simulation of 300 Megaparsecs (see Sect. \ref{section_Sims} below). The minimum mass resolved in the simulation at $z = 1.57$ is $M_{\rm min,sim} = 1.14\times 10^{11}$ and the maximum mass is $M_{\rm max,sim}=8.21\times 10^{13}$.}  
\label{DTcurves}
\end{figure}

It is interesting to show how the free-free signal depends on the redshift and the mass of the haloes. In Fig. \ref{DTcurves} we show the free-free signal for different mass intervals. In each interval, we compute the mean free-free distortion as a function of redshift. The smaller but more abundant haloes give a larger signal. Also, as we show earlier, smaller haloes have more or less the same abundance at all redshifts and hence their average free-free contribution shows a slow dependence with redshift. Note how the simulation predicts significantly less average signal than the analytical model. As we will see later, this is a direct consequence of the lack of resolution in the simulation that is not able to capture the contribution from the smallest haloes. 

This prediction, however, should be taken with care since on one hand we assumed that haloes remain ionized at all times and the temperature of the gas corresponds to the virial temperature of the halo. In small systems, the cooling time is short and the gas can cool down significantly, become neutral and form stars. Our assumptions are only valid for the most massive haloes  (groups and clusters) and the model predictions are robust only in that regime. For these objects, the average free-free distortion is of the order of a few to several tens of $\mu$K at 1 GHz. Also, the plot shows the average free-free signal obtained from simulations (see below) as a function of redshift (stars). As we will see later, the smaller range of halo masses of the simulation predicts a smaller average free-free signal.

\begin{figure}
\begin{center}
\includegraphics[width=8.0cm]{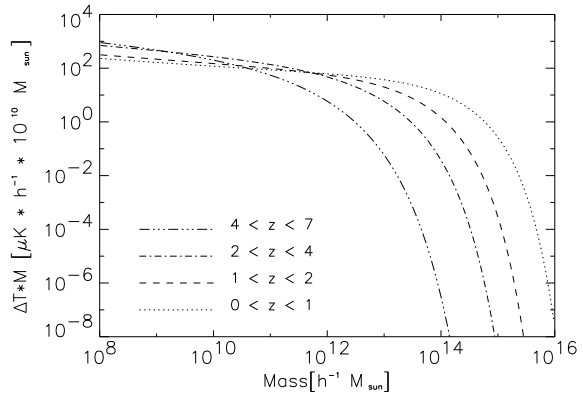}
\end{center}
\caption{Free-free signal as a function of the halo mass and for different redshift intervals.}
\label{DTvsMass}
\end{figure}

In Fig. \ref{DTvsMass} we show more explicitly the dependency of the average free-free distortion with the mass range but for different redshift intervals. Again, smaller haloes contribute more to the average signal than massive ones at all redshifts. This result is strengthened by the the model proposed in \cite{Miniati_et_al_2004}. The authors, referring to the component of the UV luminosity produced by the thermal emission from gas accreting on to dark matter haloes, calculate that the most contribution is produced by haloes with temperatures between $10^6$ K and a few $\times 10^7$ K, corresponding to masses $10^{11}-10^{13}$ solar masses.\\     

It is important to note, though, that cooling and star formation play a
critical role in determining the actual contribution of galaxy-sized haloes
($M < 10^{12} h^{-1}M_{\odot}$) to the temperature distortion of the CMB.
On the one hand, the temperature of the ionized gas will be around $10^4$ K, much
lower than predicted by Eq. (\ref{Tscalinglaw}), and its density will be considerably
higher than predicted by the $\beta$-model. The combined action of both effects can boost the expected free-free signal
by a large factor. On the other hand, a significant fraction of the gas will be 
transformed into stars and most of the interstellar medium will be in
neutral rather than ionized form, and therefore it will not emit any
bremsstrahlung radiation. The net effect is difficult to quantify, and \cite{Peng_Oh_1999} has resorted to a phenomenological parameter describing 
the fraction of {\it active} galaxies or, equivalently, the average ionization fraction 
of the gas. 
These works focus on the signal from small and cold systems, where UV radiation from stars and quasars ionizes the surrounding neutral hydrogen and creates expanding bubbles of ionized plasma. 

Gas cooling, star formation, and feedback processes determine the amount of ionized gas, its characteristic density, its temperature, and thus the total bremsstrahlung luminosity. %%In the present paper we discuss in more detail the regime of hotter, more massive objects, where the gas is collisionally rather than photo-ionized. Although these systems contribute only a small fraction of the overall cosmological signal, the physics involved is much simpler, and therefore the predicted signal constitutes a very robust lower limit. 
We discuss in more detail the regime of hotter, more massive objects, where the gas is heated collisionally rather than photo-ionized. These systems contribute only to a small fraction of the overall cosmological signal. In this work, we provide robust lower limits for the signal produced by massive objects, based on a simply physical modeling. A more detailed treatment of cooling and photo-ionization of the interstellar medium will be addressed in a future work.

%%%%%%%%%%%%%%%%%%%%%%%%%%%%%%%%%%%%%%%%%%%%%%%%%
\section{N-body simulations}\label{section_Sims}
%%%%%%%%%%%%%%%%%%%%%%%%%%%%%%%%%%%%%%%%%%%%%%%%%

In the previous section we have shown how the average free-free signal from haloes depends on their redshift and mass distributions. We also discussed how these predictions depend on the internal gas distribution inside the haloes. In this section we use numerical simulations to compute the free-free signal. Through N-body simulations we can obtain the distribution of the electron density, its temperature and ultimately the free-free effect which can be projected into sky maps.\\

We use the \gadget code (\citealp{Springel_Gadget2}). The code is a combination of a Particle Mesh Refinement algorithm and the TreeSPH method by Hernquist and Katz (\citealp{Hernquist_Katz_1989}). For the cosmological parameters we use the concordance model: $\Omega_{\rm \Lambda} = 0.73$, $\Omega_{\rm M} = 0.27$, $\Omega_{\rm b}=0.039$, $\Omega_{\rm K} = 0$, $\sigma_8 = 0.79$, $h = H_0 / (100 \mbox{ Km s}^{-1} \mbox{Mpc}^{-1}) = 0.72$ where $\sigma_8$ is the RMS mass fluctuation on a sphere of a radius of 8 Mpc.

We create the initial conditions at redshift z=49 with  the code 2LPT (\citealp{crocce-2006},) based on a second-order Lagrangian perturbation theory. The initial condition is evolved with \gadget from $z = 49$ until $z=0$. For the main simulation, we use a cosmological volume with $512^3$ particles of dark matter and $512^3$ particles of gas distributed in a box size of $(300h^{-1} \mbox{ Mpc })^3$. The force smoothing parameters has been set to 1/30 of the inter particle distance, and corresponds to 20 Kpc for the $300 h^{-1}$ Mpc simulation.
 
The outputs (or snapshots using the \gadget terminology) of the $300 h^{-1} \mbox{Mpc}$ box were chosen at redshifts for which the comoving distance between both ends of the box would overlap between consecutive redshifts. Each snapshot is analyzed independently from the others. We assume that the Universe is fully ionized below z=7 and we concentrate on this regime. The masses for the dark matter and gas particles are $M_{\rm DM} = 1.165\times 10^{10} h^{-1}M_\odot$ and $M_{\rm gas} = 0.17\times 10^{10} h^{-1}M_\odot$ respectively.
 
The minimum and maximum masses of the structures found in our simulation depend on the simulated volume, particle mass, and of course the redshift. As discussed earlier, the free-free signal has a wide dynamical range in mass.
 
The choice of the comoving volume of the simulation box is important: on one hand we want to have the largest possible box so we can include more massive haloes, on the other hand, the small structures have a very significant impact on the average free-free signal and is also important to capture the small scale signal.
 
To address the issue of resolution in the N-body simulation we make a different simulation (with the same cosmology) but with higher resolution. The use of different box sizes and resolutions is useful to study a wider range of masses (or resolutions) with N-body simulations (see for instance \citealp{Refregier_Teyssier_2002, Trenti_Stiavelli_2008}).
 
The box size of the second simulation is $(50 h^{-1}$ Mpc)$^3$ (that is $6^3$ times smaller in volume). We maintain the same number of particles ($512^3$ for dark matter and $512^3$ for gas). The resulting dark matter and gas particle masses are $M_{\rm DM} = 6.1 \times 10^7 h^{-1}M_\odot$ and $M_{\rm gas} = 0.873 \times 10^7h^{-1} M_\odot$ respectively. 
The masses of the particles are proportional to the volume of the simulation boxes divided by the number of particles, that is, since the number of particles is the same in both simulations, the particle masses are $6^3$ times larger in the 300 $h^ {-1}$ Mpc box than in the 50 $h^{-1}$ Mpc one.\\

Our simulations do not include cooling nor radiative transfer. In a future work we plan to include these mechanisms and improve the predictions. We also plan to extend the redshift range into the reionization period. For the present work, our intention is to explore the redshift range $0<z<7$ and focus on the most massive haloes for which the above effects are not so relevant. 

%%%%%%%%%%%%%%%%%%%%%%%%%%%%%%%%%%%%%%%%%%%%%%%%%%%%%%%%%%%
\subsection{Range of halo masses in the N-body simulation}
%%%%%%%%%%%%%%%%%%%%%%%%%%%%%%%%%%%%%%%%%%%%%%%%%%%%%%%%%%%
%Computing the masses of haloes from an N-body simulation is not a trivial task. Several algorithms have been used in the past for this purpose. The Friends-of-friends (FoF) algorithm was developed in 1983, by Geller and Huchra (see \citealp{Fof}). Modifications to the FoF have been proposed through the years by \cite{Fof2, Fof3, Fof4}. Other interesting finding techniques are the VoBoZ (\citealp{voboz}) and HSF (\citealp{HSF}) algorithms.\\
%More recently, a new method has been developed based on the Adaptive Mesh Refinement (AMR). The original code was proposed by \cite{mlapm} with the name of {\it multi-level adaptive particle mesh} (MLAPM). An application and improvement can be find in \cite{Gill_et_al_2004} (MHF). \\
%For our purpose, we need a halo finder codes compatible with the \gadget snapshots. 

We use a halo finder to map the distribution of haloes in mass and to associate each simulations with a proper free-free emissivity mass range (Fig. \ref{DTcurves} and Fig. \ref{DTvsMass}).
   
In order to identify haloes and subhaloes in our simulations we have run
the MPI+OpenMP hybrid halo finder AHF\footnote{AMIGA halo finder, to be
downloaded freely from \url{http://www.popia.ft.uam.es/AMIGA}}. ~A
detailed description of AHF is given in the code description paper
(\citealp{AHF}). We provide a brief summary of the mode of operation. By virtue of
the adaptive mesh hierarchy employed to sample the density field, AHF
locates overdensities as prospective halo centers.~The local potential
minima are computed for each of these density peaks and, treating the
prospective halo in isolation, the gravitationally bound particles are
determined.~Only peaks with at least 20 bound particles are considered
as haloes and retained for further analysis. ~For each halo, we compute
the virial radius $r_{\rm vir}$, that is the radius $r$ at which the
density $M(< r)=(4\pi r^3 / 3)$ drops below $\Delta_{\rm vir}
\bar{\rho}$.  Here $\bar{\rho}$ is the cosmological
background density. ~The threshold $\Delta_{\rm vir}$ is computed
using the spherical top-hat collapse model and is a function of both
cosmological model and time. 
 
%%%The result is shown in Fig.\ref{massfunction300Mpc}, where it is compared with the mass function of \citealp{Sheth_Mo_Tormen}. 
%\begin{figure}
%\begin{center}
%\includegraphics[width=8.8cm]{compare_ST.ps}
%\end{center}
%\caption{Theoretical integrated mass function (solid line) compared with the integrated mass functions derived from our simulations (50 $h^{-1}$ Mpc dotted line and 300 $h^{-1}$ Mpc dashed line). In all cases the redshift is $z=1.57$.
%\label{massfunction300Mpc}
%\end{figure}
Applying the AHF to the 300 Mpc simulation, we have found that the mass of the inside haloes only span between $M \approx 10^{11}h^{-1}M_\odot$ and $M \approx 10^{14}h^{-1}M_\odot$. Low mass haloes are not present in the simulation due to the resolution. On the high end mass, the limited volume of our simulation prevents us from having the most massive clusters in our simulation.

%%%%%%%%%%%%%%%%%%%%%%%%%%%%%%%%%%%%%%%%%%%%%%%%%%%%%
\subsection{Average free-free from the simulation}
%%%%%%%%%%%%%%%%%%%%%%%%%%%%%%%%%%%%%%%%%%%%%%%%%%%%%

For each gas particle in the volume, we compute the free-free luminosity assuming that the electron density is approximately constant over the volume of the particle. Then, the integral of the squared of the electron density can be computed as: 
\begin{equation}
\int_V n_{\rm{e}}^2 dV \approx n_{\rm{e}} \int_V n_{\rm{e}} dV = n_{\rm{e}} \frac{M_{\rm gas}}{\mu m_{\rm p}}.
\label{particlecontribution}
\end{equation}

The gas density, $n_{\rm{e}}$, at the position of the particle is extracted from \gadget and then transformed into convenient cm$^{-3}$ units.  

Eq. (\ref{particlecontribution}) is used to compute the particle luminosity from Eq. (\ref{ff_c}). The particle luminosity is transformed into particle flux given the luminosity distance, $D_{\rm L}$, from the particle at redshift $z$ to the observer (at $z=0$). 
\begin{equation}
S_{\rm ff}(Jy) = \frac{e_{\rm{\nu}}}{4\pi D_{\rm L}^2}.
\label{part_brightness}
\end{equation}

The internal energy is given by \gadget in units of $[\mbox{km sec}^{-1} ]^2$ which is converted into K with the factor:
\begin{equation}
C_{\rm K} = 10^6 (\gamma-1) \mu \frac{m_{\rm p}}{k_{\rm b}},
\label{en_conversion}
\end{equation}
where $\gamma = 5/3$ is the adiabatic index for a monoatomic ideal gas. The scale factor $10^6$ accounts for \gadget's internal units, and $k_{\rm b}$ is the Boltzmann constant. 

After the flux per particle is computed, the fluxes are projected along the line of sight into a pixelized 2D map.~Since the apparent angular size of each box depends on the redshift, we have to restrict our analysis to the smallest field of view that in our case corresponds to the apparent size of the most distant box (about 3 degrees for the 300 $h^{-1}$ Mpc box). Because we want to compare the distortion that our model induces on the CMB temperature as a function of frequency, we extract a mean flux, $\bar{S}_{\rm ff}$, from all the projected maps. The resulting mean brightness is converted into temperature in K (antenna or thermodynamic, since we are considering low frequencies):
\begin{equation}
\Delta T(\nu) = \frac{c \bar{S}_{\rm ff}}{2 k_{\rm b} \nu^2}
\label{antenna_conversion}
\end{equation}

\begin{figure}
\begin{center}
\includegraphics[width=8.0cm]{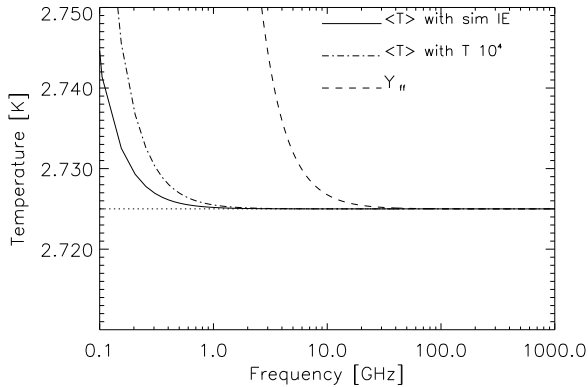}
\end{center}
\caption{Free-free emission distortion from the 300 $h^{-1}$ Mpc simulation. The emission is computed from 20 snapshots within the interval $0 < z < 7$. The field of view covers 2.7 degrees. The solid line shows the distortion in the case where the emissivity has been computed with the temperature derived from the simulation. The dash-dotted line corresponds to the case where the temperature for all particles has been fixed to $10^4$ K. The dashed line shows the observational upper limit constraints (95\% CL) from \citep{Bersanelli_et_al_1994}}
\label{distortion_300Mpc}
\end{figure}

In Fig. \ref{distortion_300Mpc} we show the result obtained from Eq. (\ref{antenna_conversion}).
% (where $\bar{S}_{ff}$ was computed at $\nu = 1$ GHz from the simulations and we neglected the small effect of the frequency dependence of the gaunt factor).
%We plot the actual constrain of $Y_{\rm ff}$ ({\it dashed line}), the distortion produced by the free-free emission is physically related to the Comptonization distortion.
We plot, based on the actual constraint of $Y_{\rm ff}$ ({\it dashed line}), the corresponding upper limit distortion produced by the free-free emission over the CMB temperature.
%We compare these results with the constrain on the optical depth due to Compton scattering $y$ (see the introduction). 
The solid line shows the mean temperature distortion of the projected map in the sky from the simulation while the dot-dashed line refers to the assumption of a constant temperature for all particles of $T=10^4$ K (see \citealp{Peng_Oh_1999}). Because in Eq. (\ref{ff_c}) the strength of the signal depends inversely on the temperature, the lower the temperature of the gas, the higher the signal.

From Fig. \ref{distortion_300Mpc}, we conclude that the average free-free distortion predicted from our 300 $h^{-1}$ Mpc N-Body is well below the current observational constraint ({\it dashed} line). A much lower temperature for the gas ($10^4$ K) in Eq. (\ref{ff_c}) does not change the effect too much showing the anticipated weak dependency of the free-free distortion with the temperature. 

The signal from the simulation is also significantly smaller than the value predicted using the analytical model.~As we will see below, the most likely reason for this is the fact that the simulation does not include the small mass haloes that give most of the signal in the analytical case. A limiting factor of the N-body simulations is that by construction there is a minimum mass for the haloes. This can have a large impact on the predicted average signal as smaller haloes are expected to be much more numerous than massive ones and they can boost the average signal. In the next subsection we explore the range of masses present in the simulation.

%%%%%%%%%%%%%%%%%%%%%%%%%%%%%%%%%%%%%%%%%%%%
\subsection{Dependency with the resolution}
%%%%%%%%%%%%%%%%%%%%%%%%%%%%%%%%%%%%%%%%%%%%

In the previous sections, we have shown the results obtained with the 300 Mpc simulation. In this section we compare the results obtained with the $50 h^{-1}$ simulation that has a much higher resolution.
   
When we compare the mass functions, we find that, as expected, the 50 $h^{-1}$ Mpc box contains less massive haloes, but it has many more small haloes. 
%(see figure \ref{massfunction300Mpc}). 
A halo must contain of the order of 20 particles to be considered a halo. Therefore the minimum mass of a halo depends on the resolution of the simulation. On the other hand, the maximum mass of a halo depends on the volume of the simulation. Large haloes are truncated by the boundary conditions of the simulation that suppress the power on scales larger than the box side.~In other words, there is a minimum $k$-mode in the Fourier modes which is directly related with the dimension of the box. 
%%%The mass function of the 50 pc box is shown in figure \ref{massfunction300Mpc}. Both mass functions (from the 50 and 300 $h^{-1}$Mpc simulations) agree well in the regime $10^{12} h^{-1}M_\odot<M<10^{13} h^{-1}M_\odot$.  

We compare the average free-free effect in the 300 $h^{-1}$ Mpc and 50 $h^{-1}$ Mpc boxes. Since the 50 $h^{-1}$ Mpc box is $6^3$ times smaller in volume than the 300 $h^{-1}$ Mpc one, we renormalize the average free-free to the same volume. As expected, due to the presence of smaller haloes in the 50 $h^{-1}$ Mpc simulation, the smaller box produces a larger free-free signal. Considering a slice of 50 $h^{-1}$ Mpc at redshift 1.57 in both cases; in the 300 $h^{-1}$ Mpc box the average $\Delta T$ is $\Delta T \approx 10^{-6}$ K at 1 GHz while in the 50 $h^{-1}$Mpc box $\Delta T \approx 5 \times 10^{-6}$ K also at 1 GHz. This is a factor 5 more signal in the higher resolution case. As shown earlier, this extra signal comes from the lower mass haloes although we should keep in mind that in our model we are not including neither radiative cooling nor partial ionization of the low mass haloes. These effects compensate each other partially (in terms of the free-free signal) but they will change the amount of free-free predicted by our model (again, in the low mass haloes more than in the massive ones).

%%%%%%%%%%%%%%%%%%%%%%%%%%%%%%%%%%%%%%%%%%%%%%%%%%%%%%%%%%%%%%%%%%%%%%%%%%%%%%%%%%%
\section{Free-free from a single massive halo. A new window for cluster science ?}
%%%%%%%%%%%%%%%%%%%%%%%%%%%%%%%%%%%%%%%%%%%%%%%%%%%%%%%%%%%%%%%%%%%%%%%%%%%%%%%%%%%

In the previous section we have shown how the average contribution of the massive haloes (groups and clusters) to the mean free-free signal is significantly smaller than the contribution from the smaller but more numerous low mass haloes. In this section we explore the signal of an individual halo comparing the prediction from the $\beta$-model with the result obtained from the numerical simulation.\\ 

Using the high resolution simulation (50 $h^{-1}$ Mpc box, $M_{\rm gas} = 0.8\times 10^7 h^{-1}M_\odot$ per particle) presented in Sect. 4, we extract the most massive cluster from it in order to compare its free-free flux with an analytical model. The redshift of the simulation is $z=1.6$ but the same conclusions can be extracted at other redshifts. It is however interesting to explore the high redshift regime since the free-free signal could be potentially useful to detect clusters in their earlier stages of formation and before the gas is too hot to be seen through X-rays. In this sense, the free-free emission could extend the actual X-ray science in clusters to the range of the radio waves. Similarly, the same cluster could be seen through the SZ effect but its detection will be harder if the cluster is not hot enough. On the contrary, a lower temperature in the cluster makes the free-free signal stronger.
  
\begin{figure}
\begin{center}
\includegraphics[width=8.0cm]{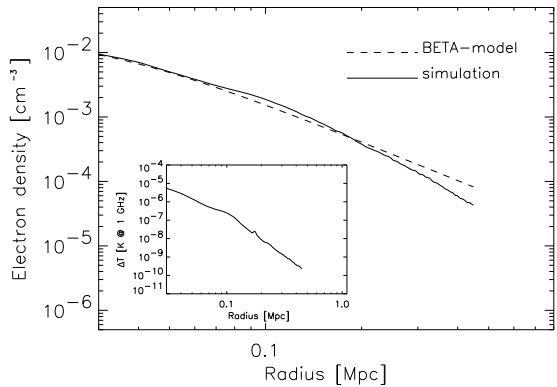}
\end{center}
\caption{Density profile extracted from the most massive halo in the 50 $h^{-1}$ Mpc simulation box at $z=1.6$. The solid line shows the average electron density profile in concentric shells. For comparison, a $\beta$-model is shown (dashed line). The model corresponds to a core radius $R_{\rm c} \sim 40$ kpc and  $p = 1/14$. The smaller box shows the temperature distortion produced by this halo as a function of the radius and at $\nu = 1$ GHz. The maximum distortion is $\Delta T / T_{\rm CMB} \approx 10^{-6}$.} 
\label{HALO_BETA_profile}
\end{figure}

The halo boundaries are defined as  $R_{\rm vir}$ or the radius where the over-density equals 200 times the average density in the box (according to the common assumption of the virial radius $r_{200}$, adopted to our simulation). For the most massive halo in our simulation, this radius corresponds to $R_{\rm vir} = 560$ kiloparsecs (co-moving) and the corresponding total mass of the halo within this radius is $M_{\rm halo} = 9 \times 10^{12}h^{-1}M_\odot$.~This mass corresponds to a group of galaxies. We add the flux per particle  and compute the flux and temperature distortion (see Eqs. (\ref{ff_c}) and (\ref{particlecontribution})). In Fig. \ref{HALO_BETA_profile}, we show the one-dimensional profile. The solid line represents the electron density (in $\mbox{cm}^{-3}$) as a function of the radius. In order to compare this profile with a $\beta$-model, the values for $n_o$ and $R_c$ of the $\beta$-model are obtained by fitting the solid line in Fig. \ref{HALO_BETA_profile} with the analytical profile. The best fitting $\beta$-model is also shown in Fig. \ref{HALO_BETA_profile} where the core radius corresponds to 1/14 of the virial radius (or $p=14$ in the notation used above).

\begin{figure}
\begin{center}
\includegraphics[width=8.0cm]{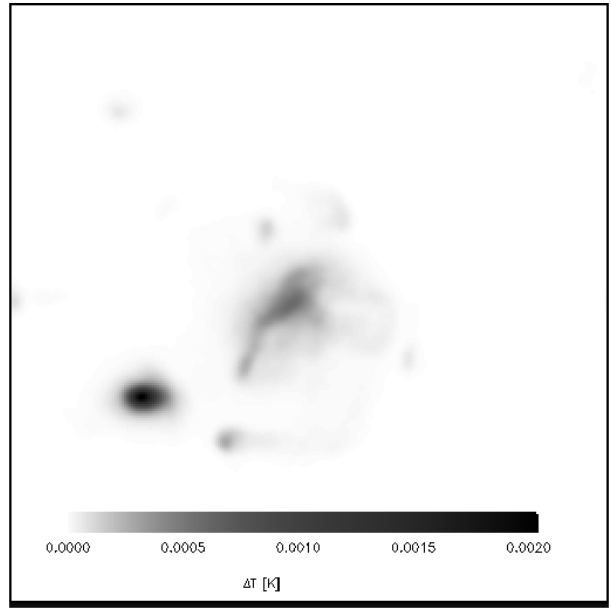}
\end{center}
\caption{Free-free distortion for a massive halo ($M = 6.6 \times 10^{14} h^{-1} M_\odot$) at redshift $z=0.15$. The greyscale shows the distortion in K and at 1 GHz. The field of view is $\approx 40'$. The total flux in this region is  $S_{\rm ff} = 2.83\times 10^{-5}$ Jy. }
\label{Map_DeltaT}
\end{figure}

We fix the temperature for the $\beta$-model case to the average over the halo particles in the simulation. This average corresponds to $T_{halo} \approx 10^6$ K. With all these ingredients, the fluxes for this halo can be derived from the simulation and the analytical model. For the simulation case we find $S_{\rm ff} = 3.67\times 10^{-9}$ Jy, while the $\beta$-model predicts a larger flux $S_{\rm ff} = 2.80 \times 10^{-8}$ Jy (a factor 7.5 larger). 
In terms of $\Delta T$, the maximum temperature distortion is about a few $\mu$K (at 1 GHz) at the center of the cluster, that is, within reach of future planed experiments like SKA. According to \cite{Burigana_et_al_2004}, SKA could reach a sensitivity limit of 40 nJy in one hour of integration and with an angular resolution of 1 milliarcsecond in the 4-20 GHz band. More massive and denser clusters would produce an even stronger signal making the study of free-free emission in clusters at radio frequencies an interesting and useful way to study the intracluster medium. In Fig. \ref{Map_DeltaT} we show a map of the free-free signal at 1 GHz in an area containing a more massive cluster at redshift $z=0.15$ extracted from the 300 Mpc simulation.~In this case, the free-free distortion is of the order of 1 mK in the cluster regions. At higher frequencies, the temperature distortion decreases as $\nu^{-2}$. That is, at 30 GHz, the temperature distortion would be of the order of 1 $\mu$K.     

%%%%%%%%%%%%%%%%%%%%%
\section{Discussion}
%%%%%%%%%%%%%%%%%%%%%

Our results show that there is a significant free-free signal at all redshifts up to the 
time of reionization. Our predictions are based on analytical models and they are compared with N-body simulations. Some assumptions made in our model need to be improved, like, for instance, the fact that all low mass haloes remain ionized at all times.

Another important improvement is to substitute the $\beta$-model (in the analytical calculations) by a more accurate description of the gas in massive haloes. In particular, the model of \cite{ascasibar_diego_2008} assumes a steeper and non-isothermal profile for the gas distribution that could boost the free-free signal. This model is in better agreement with high resolution X-ray profiles in galaxy clusters (\citealp{ascasibar_diego_2008, Sanderson_Ponman_2009}) and with the SZ effect (\citealp{Diego_Ascasibar_2008, Diego_Partridge_2009}) than the $\beta$-model.

Another issue that need to be addressed in the future is the fact that the free-free effect is significant for a wide range of halo masses. This fact, combined with the high range of redshifts, makes the computation of the free-free from simulations a very demanding task from the computational point of view.

%%%%%%%%%%%%%%%%%%%%%%%%%%%%%%%%%%%%%%%%%
\subsection{Comparison with earlier results}
%%%%%%%%%%%%%%%%%%%%%%%%%%%%%%%%%%%%%%%%%

It is interesting to compare our results (based on numerical and analytical analysis) with those found in the literature that use only analytical methods (\citealp{Haiman_Loeb_1997,Peng_Oh_1999, Cooray_Furlanetto}). The main difference between our analysis and previous ones is that we have focused on the better understood regime at lower redshifts and higher masses. \cite{Cooray_Furlanetto} shows how the free-free signal has the maximum contribution at redshift $z \leq 3$. The free-free signal is integrated from the beginning of the reionization ($z \sim 12$ in \citealp{Peng_Oh_1999} and $z \sim 13$ in \citealp{Cooray_Furlanetto}) until present while we consider only a redshift range ($0<z<7$) in which the Universe can be considered as fully ionized (on large scales). Also, in this work we focus more on massive haloes which are the ones that can be considered as fully ionized at all times (for $z<7$). In earlier works, only low mass haloes where considered in the calculations of the free-free signal. The modeling of the low mass haloes is more difficult since they are more sensitive to non-linear phenomena. In a future work, we will extend our analysis to higher redshifts to include the transition between a neutral and a ionized Universe and a more careful modeling of the low mass haloes.

In \cite{Peng_Oh_1999} (see also \citealp{Oh_Mack}), a model is proposed for the ionizing sources. The model includes the production rates of recombination line photons $\dot{N}_{\rm recomb}$ and ionizing photons $\dot{N}_{\rm ion}$. It makes a clear distinction between virialized (collapsed) structure that undergo a starburst phase and a diffuse gas that is constantly being re-ionized. A halo mass function is used to compute the number of active haloes (or haloes with a starburst, and ionizing UV flux) and the duration of the starburst is set to a constant interval of $t_0 = 10^7$ years. Our model is, instead, much more simplistic and assumes that all haloes are fully ionized. This assumption certainly fails in the low mass halo regime. 

In \cite{Peng_Oh_1999},the emissivity $\epsilon_{\rm \nu}$ is computed combining an expression for the luminosity of the haloes $L_{\rm \nu}(M,z)$ and the rate formation of ionizing photons. In this model, the temperature is fixed to $10^4$ K. % uses an expression for the luminosity of the haloes $L_{\nu}(M,z)$, combined with Eq. (\ref{ff_c}) and the rate formation of ionizing photons as a function of the mass of the halo to compute the emissivity from the haloes. The temperature of the haloes is also fixed to $\sim 10^4$ K.
 
In our case, we used a combination of a $\beta$-model plus the mass function combined with a scaling law for the temperature in the analytical case. In the N-body simulation, no assumptions are made about the gas profile or its temperature since these values are extracted directly from the simulation. Recent models (\citealp{Fardal_et_al}) show that the gas is seldom heated up to the virial temperature in systems with  $T < 10^6$ K. Instead they are accreted in {\it cold flows} (with $T \sim 10^4$ K). The cold flow mechanism is not implemented in our N-body simulations resulting in smaller free-free signal from the smallest haloes. 

The N-body simulation includes the contribution from both, compact haloes and diffuse IGM. In the work by \cite{Peng_Oh_1999}, (and later by \citealp{Oh_Mack}), a clear distinction is made between the contribution from small ionized haloes (that remain ionized for a limited amount of time before becoming neutral again) and the diffuse IGM. The authors introduce a cutoff flux $S_{\rm c}$ corresponding to the minimum mass able to be ionized and with a temperature of $T_e = 10^4$ K. The minimum mass for the ionized haloes with this temperature evolves with redshift as $M_{\ast}\sim 10^8 (1+z/10)^{-3/2}$ $h^{-1}M_{\odot}$. In our case, the temperature is derived from the simulation and is, in general, larger than the temperature used in \cite{Peng_Oh_1999} and \cite{Cooray_Furlanetto}. As a consequence, our higher temperatures will predict a lower free-free signal from haloes. As we mentioned earlier, in a future work we plan to include mechanisms such as cooling that would reduce the temperature of the haloes (and hence would boost the free-free signal). 

The distortion over the CMB temperature from haloes derived by \cite{Peng_Oh_1999} is $\Delta T_{\rm ff} = 3.4 \times 10^{-3}$ K at 2 GHz. This result was obtained with no cutoff in the flux of point sources ($S_{\rm c} = 0$). On the other hand, an estimation of the flux coming from the diffuse IGM renders a much smaller temperature distortion ($\Delta T_{\rm ff} = 6.0 \times 10^{-6}$ K at 2 GHz) a result later confirmed by \cite{Cooray_Furlanetto}. 
\cite{Cooray_Furlanetto} obtain a value of $\Delta T_{\rm ff} \approx 5.0 \times 10^{-3}$ K for the halo contribution also at 2 GHz. 

Comparing these numbers with our analytical predictions (see Fig. \ref{FFmodel}), we obtain a lower signal at 2 GHz when the temperature of the haloes is computed with the scaling law Eq. (\ref{Tscalinglaw}) ($\approx 1.73 \times 10^{-3}$ K). Fixing the temperature to $\sim 10^4$ K the results agree better ($\approx 7 \times 10^{-3}$ K).

%%%%%%%%%%%%%%%%%%%%%
\section{Conclusions}
%%%%%%%%%%%%%%%%%%%%%

As can be seen in Fig. \ref{DTcurves}, the mean free-free signal is larger for lower mass haloes. On the other hand, large mass haloes have larger individual free-free fluxes but they are much less abundant. Consequently their contribution to the mean signal decreases quickly with increasing mass.  

An interesting result was shown in Fig. \ref{DTvsMass} where the explicit dependency of the average temperature distortion with the mass is shown for different redshifts intervals. From this plot, it is clear that the average signal is most sensitive to halo masses smaller than $10^{12}$ solar masses. Also, from the same figure we can conclude that haloes contribute to the average free-free signal at all redshifts up to the reionization time. 

Even though groups and clusters are expected to contribute less than less massive haloes to the average signal, it should be possible to detect clusters through their free-free signal on a one by one basis opening the door for interesting studies of the intra-cluster medium at radio wavelengths. In this line, \cite{Cooray_Furlanetto} discussed the possibility of detecting the signal from clumps of IGM with ARCADE. In \cite{Burigana_et_al_2004} the model by \cite{Peng_Oh_1999} for unresolved free-free emitters has been exploited arriving to the indication that the SKA project will be able to detect them with deep exposures.%estimate the emission from resolved free-free emitters. They conclude that the SKA project will be able to detect them with deep exposure. They predict $\sim 10^4$ detections from individual free-free emission sources with $z > 5$ in 1 square degree above a source detection threshold of 70 nJy. 

Future experiments might focus on the detectability of individual groups and/or clusters through their free-free signature. This signal can be combined with others (SZ, X-rays) in multiwavelength studies.

\section*{Acknowledgments}
We would like to thank the referee for useful comments that helped to improve this paper.

We acknowledge partial financial support from the Ministerio de Ciencia e Innovaci\'on project AYA2007-68058-C03-02.~PPP acknowledges support from the Spanish Ministerio de Educaci\'on y Ciencia and CSIC for an I3P grant.~PPP also acknowledges support from the program of ``Estancias Breves'', grants 22007ESTI3P-00493 and 009ESTCSIC-01386.~PPP also thanks the IASF/INAF, Istituto di Astrofisica Spaziale e Fisica Cosmica, Istituto Nazionale di Astrofisica, and the University of Pennsylvania for their hospitality during parts of this research.~CB acknowledges partial support for this work by the ASI/INAF Agreement I/072/09/0 for the Planck LFI Activity of Phase E2 and the ASI contract I/016/07/0 COFIS.~RKS acknowledgments for the support of grant NFS-AST 0908241.~SRK acknowledges support by the MICINN under the Consolider-Ingenio, SyeC project CSD-2007-00050.~YA acknowledgs financial support from project AYA2007-67965-C03-03.

Some of the results of this work have been obtained using the Altamira supercomputer at IFCA which is part of the Red Espa\~nola de Supercomputaci\'on.~Finally, the authors would like to thank Peng Oh and Asantha Cooray for useful comments.

\bibliographystyle{mn2e}
\bibliography{FF_bib}

\appendix
\section{Computing the mass function}
\label{appendix_a}
In ST, the mass function is given in terms of the factor $\nu \equiv [\delta_c/\sigma_M]^2$, where $\delta_c$ is the overdensity contrast required for the perturbation to collapse, and $\sigma_M$ is the rms fluctuation in the mass scale $M$. The function
\begin{equation}
\nu f(\nu) = M^2 \frac{n(M,z)}{\bar{\rho}}\frac{d\log M}{d\log \nu}
\label{univ_scale}
\end{equation}
behaves like an almost  universal function with respect to changes in redshift (\citealp{Tinker_MF}). The quantity $\bar{\rho} = 2.775 \times 10^{11} \Omega_{\rm M} h^2 M_\odot \mbox{ Mpc}^{-3}$ is today's average matter density.

This mass function accounts for the fact that the gravitational collapse of a halo is not exactly spherical but rather if follows a triaxial model. For a given cosmological model, the evolution of an ellipsoidal perturbation is determined by three parameters, namely the eigenvalues of the deformation tensor. These are the ellipticity $e$, the prolateness $p$ and the density parameter $\delta$. In their model, the collapse is traced independently over the three orthogonal axes and the virialization of the halo is defined as the time when it collapses along the three axes. Since each axis collapses independently from the others, collapse along each axis is frozen once it has shrunk by some critical factor.
 
The term $\nu f(\nu)$ in Eq. (\ref{univ_scale}) is parametrized in the ST formalism as:
\begin{equation}
\nu f(\nu) = A \left ( 1+ \frac{1}{\nu'^p} \right ) \left ( \frac{\nu'}{2}\right ) ^{1/2} \frac{e^{-\nu'/2}}{\sqrt{\pi}},
\label{sheth-tormen}
\end{equation}
where $\nu' = a\nu$, $a = 0.707$, and $p =0.3$. In the standard Press and Schechter mass function (\citealp{press_schechter}) $p=0$. $A\approx 0.3220$ is the normalization factor given by the constraint that the integral of $f(\nu)$ in the whole $\nu$ range must be equal to 1. For comparison, in the original formalism of Press and Schechter, the normalization is 1/2.

%\begin{eqnarray}  %P-S formalism
%\frac{dN(M,z)}{dzdM} = & \sqrt{\frac{2}{\pi}} \frac{\bar{\rho}}{M^2}\frac{\delta_{c0}(z)}{\sigma_M} \left | \frac{d\log \sigma_M}{d\log M} \right | \times & \nonumber \\
 %& \exp\left [\frac{1}{\sqrt{2}}\frac{-\delta_{c0}^2(z)}{2\sigma_M^2} \right ] & ,
%\end{eqnarray}
%where $\bar{\rho} = \Omega_m \cdot 2.775 \cdot 10^{11}\cdot h^2 M_\odot \mbox{ Mpc}^{-3}$ is the average matter density today ($\Omega_m$ the matter density parameter) and $\delta_{c0}(z)$ is the over-density of a collapsed sphere at redshift $z$ in linear theory. 

In Eq. (\ref{sheth-tormen}), the over-density parameter $\delta_c$ can be estimated, given a cosmological model, using the linear growth parameter $D(z)$ and $\delta_c(z)$ (see for example \citealp{Mathiesen_Evrard}):
\begin{equation}
\delta_{\rm c} = \frac{D_0)}{D(z)} \delta_{\rm c}(z).
\end{equation}

Similarly one could have considered $\delta_c(z)=$ const and $\sigma_{\rm M}(z) = \sigma_{\rm M} D(z)/D_0$ with the same results. 
The mass function, $n(M,z)$, can be easily derived from Eq. (\ref{univ_scale}) and Eq. (\ref{sheth-tormen}). \\

We assume a flat $\Lambda$CDM model ($\Omega_{\rm M} + \Omega_{\rm \Lambda} =1$). In this case $\delta_{\rm c}(z) = 1.6866[1 + 0.01256 \log_{10} \Omega_{\rm M}(z)]$. The linear growth factor is given by \citep{Peebles_1980}
\begin{equation}
D(x) = \frac{\sqrt{x^3+2}}{x^{3/2}} \int^x_0 x'^{3/2}(x'^3 +2)^{-3/2} dx',
\end{equation}
where $x = a/[(1-\Omega_{\rm \Lambda})/(2\Omega_{\rm \Lambda})]^{1/3}$. In Eq. (\ref{sheth-tormen}) $\sigma_{\rm M}$ is the RMS fluctuation on the mass scale $M$:
\begin{equation}
\sigma_{\rm M}^2 = \frac{1}{2 \pi} \int_0^\infty dk k^2P(k) W^2(kR).
\label{RMS}
\end{equation}

The window function $W(kR)$ is a top hat function in Fourier space
\begin{equation}
W(kR) = \frac{3(\sin(kR)-kR\cos(kR))}{(kR)^3}
\end{equation}
with $R$ defined by  $M=\frac{4}{3}\bar{\rho}R^3$.
The power spectrum $P(k)$ can be parametrized as
\begin{equation}
P(k) = A k^n T^2(k),
\label{powerspectrum}
\end{equation}
where $A$ is the amplitude and $T(k)$ the transfer function.~This choice for the amplitude makes it possible to introduce it in Eq. (\ref{RMS}) with $R=8h^{-1} \mbox{ Mpc }$ to obtain the value of $\sigma_8 = 0.8$, while the index for primordial power spectrum is set to $n=1$ (both values are from the fifth year of WMAP analysis, \citealp{Dunkley_et_al_2009}). For the transfer function we use the expression given in \cite{Bardeen_et_al_1986}
\begin{eqnarray}
T(k) = & \frac{\ln (1 + 2.34q)}{2.34q} \left [ 1 + 3.89q+(16.1q)^2+(5.46q)^3 \right. &\\ \nonumber
& \left. +(6.71q)^4 \right ]^{-\frac{1}{4}},& \\ \nonumber
\label{transferfunction}
\end{eqnarray}
where $q = k (h^{-1}\mbox{Mpc})/\Gamma$, and $\Gamma$ is the shape parameter of the power spectrum ($\Gamma \sim \Omega_{\rm M} h$). We have compared the mass functions obtained with this transfer function and the more elaborated one in \citep{Eisenstein_Hu_1998} finding very small differences. For simplicity we use the \cite{Bardeen_et_al_1986} transfer function.

\end{document}